\documentclass[journal=jctcce, manuscript=article, layout=twocolumn]{achemso}
\usepackage{graphicx}
\usepackage{amsmath,physics,dsfont}
\usepackage{color}
\usepackage{microtype}
\usepackage{defs-private}

\definecolor{forestgreen}{rgb}{0.13, 0.55, 0.13}

\title{Efficient step-merged quantum imaginary time evolution algorithm for quantum chemistry}

\author{Niladri Gomes}
\affiliation{Ames Laboratory, U.S. Department of Energy, Ames, Iowa 50011, USA}

\author{Feng Zhang}
\affiliation{Ames Laboratory, U.S. Department of Energy, Ames, Iowa 50011, USA}

\author{Noah F. Berthusen}
\affiliation{Ames Laboratory, U.S. Department of Energy, Ames, Iowa 50011, USA}
\alsoaffiliation{Department of Electrical and Computer Engineering, Iowa State University, Ames, Iowa 50011, USA.} 
 
\author{Cai-Zhuang Wang}
\affiliation{Ames Laboratory, U.S. Department of Energy, Ames, Iowa 50011, USA}
\alsoaffiliation{Department of Physics and Astronomy, Iowa State University, Ames, Iowa 50011, USA}

\author{Kai-Ming Ho}
\affiliation{Ames Laboratory, U.S. Department of Energy, Ames, Iowa 50011, USA}
\alsoaffiliation{Department of Physics and Astronomy, Iowa State University, Ames, Iowa 50011, USA}

\author{Peter P. Orth}
\affiliation{Ames Laboratory, U.S. Department of Energy, Ames, Iowa 50011, USA}
\alsoaffiliation{Department of Physics and Astronomy, Iowa State University, Ames, Iowa 50011, USA}

\author{Yongxin Yao}
\email{ykent@iastate.edu}
\affiliation{Ames Laboratory, U.S. Department of Energy, Ames, Iowa 50011, USA}
\alsoaffiliation{Department of Physics and Astronomy, Iowa State University, Ames, Iowa 50011, USA}

\begin{document}

\begin{abstract}
    We develop a resource efficient step-merged quantum imaginary time evolution approach (smQITE) to solve for the ground state of a Hamiltonian on quantum computers. This heuristic method features a fixed shallow quantum circuit depth along the state evolution path. We use this algorithm to determine binding energy curves of a set of molecules, including H$_2$, H$_4$, H$_6$, LiH, HF, H$_2$O and BeH$_2$, and find highly accurate results. The required quantum resources of smQITE calculations can be further reduced by adopting the circuit form of the variational quantum eigensolver (VQE) technique, such as the unitary coupled cluster ansatz. We demonstrate that smQITE achieves a similar computational accuracy as VQE at the same fixed-circuit ansatz, without requiring a generally complicated high-dimensional non-convex optimization. Finally, smQITE calculations are carried out on Rigetti quantum processing units (QPUs), demonstrating that the approach is readily applicable on current noisy intermediate-scale quantum (NISQ) devices.
\end{abstract}

\maketitle
\section{Introduction}
One of the most promising near-term applications of quantum computing is to solve the electronic structure of molecules and condensed matter systems~\cite{feynman82qc, asp_ipea, rmp_qcc, cr_qcc, qa_dualbasis, hybrd_dmft, gqce}. This is because the number of binary bits required to store a general many-body state of a fermionic Hamiltonian grows exponentially with the dimension of the single-particle basis in classical computers, while quantum computers offer a natural representation of many-body states using qubits whose required number only scales linearly with the size of the single-particle basis. A many-body wave function can thus be efficiently stored in memory using qubits. The pioneering proposal of quantum phase estimation algorithm (PEA) needs $\O(1/\epsilon)$ controlled-$U$ operators and $\O(\log(1/\epsilon))$ ancillary qubits to reach an accuracy $\epsilon$, where $U$ is the time-evolution operator of a given system Hamiltonian~\cite{pea_kitaev, pea_lloyd}. This represents a very stringent requirement for the quantum resources in terms of number of qubits, gate fidelity and coherence time, which is beyond the current or near-term NISQ computing technology. While the number of ancillary qubits can be significantly reduced by adopting the recursive PEA~\cite{asp_ipea}, the general condition of deep quantum circuits in the PEA and the adiabatic state preparation (ASP) remains prohibitive for practical calculations on NISQ devices. 

A large class of algorithms adapted to NISQ hardware have been developed in recent years, to exploit the new technology in Hamiltonian simulations, or a wider set of optimization problems~\cite{qaoa, vqe, wecker2015_trotterizedsp, vqe_theory, vqe_pea_h2, qml, hardware_efficient_vqe, VQE_qcc, kUpUCCGSD, qa_dualbasis, vqe_adaptive, MayhallQubitAVQE}. The variational quantum eigensolver (VQE) represents a most promising approach to address open quantum chemistry problems using NISQ technologies~\cite{vqe, wecker2015_trotterizedsp, vqe_theory, vqe_pea_h2, hardware_efficient_vqe}. Within VQE, the state wavefunction is parameterized by a variational ansatz. The cost function, which is usually the expectation value of the system Hamiltonian with respect to the variational ansatz, can be efficiently calculated on NISQ devices with relatively shallow circuits. The variational parameters are adjusted to extremize the cost function using classical computers. The effectiveness of VQE is determined by the variational wavefunction form and the  high-dimensional classical optimization. The unitary coupled cluster ansatz with single and double excitations (UCCSD) represents a commonly used variational form, motivated by the success of the CCSD method in classical quantum chemistry calculations for systems free of multi-reference characters~\cite{ccsd, rmp_cc, GustavoUCC}. Many efforts have been devoted to improve the variational ansatz regarding the computational accuracy and variational circuit complexity~\cite{hardware_efficient_vqe, VQE_qcc, kUpUCCGSD, qa_dualbasis, vqe_adaptive, MayhallQubitAVQE, ooUCCSD}. For examples, the hardware-efficient ansatz prepares the variational state by a sequence of native two-qubit entangling gates alternating with single qubit Euler rotations to an initial state such as Hartree-Fock (HF) state~\cite{hardware_efficient_vqe}. The $k$-UpCCGSD ansatz is composed of $k$ products of generalized unitary paired double excitations and a complete set of generalized single excitations, which can be systematically improved toward exact answers~\cite{kUpUCCGSD}. The quantum approximate optimization algorithm (QAOA) provides an alternative way to construct a variational ansatz in the form of applying the system Hamiltonian and mixing Hamiltonian to a reference state~\cite{qaoa}. The variational wavefunction form has also been proposed to be dynamically optimized, which provides a compact system-dependent ansatz with systematically improvable accuracies~\cite{VQE_qcc, vqe_adaptive, MayhallQubitAVQE}. 

While the variational wavefunction form in VQE can be optimized to some extent, the number of variational parameters is deemed to grow with the system size under study. The cost function of VQE is generally non-convex in the high-dimensional parameter space, which renders the classical optimization problem susceptible to local minima and very challenging~\cite{vqe_theory}. Recently, a quantum imaginary time evolution algorithm (QITE) has been proposed as an alternative approach to determine eigenstates of an Hamiltonian on quantum computers without the complication of high-dimensional optimization~\cite{qite_chan20}. The idea originates from the classical imaginary time evolution algorithm, which is a sophisticated way to obtain Hamiltonian eigenstates using classical computers~\cite{imaginarytime, ite_Eloranta}. Within the QITE algorithm, the non-unitary imaginary time evolution operator is replaced by a unitary operator which preserves the induced variation in the quantum state. The unitary operator is uniquely determined by solving a system of linear equations and can be conveniently applied on quantum computers. The QITE method has been demonstrated by solving a set of finite spin models on quantum simulators, including a two-site Ising model and H$_2$ dimer on real quantum devices~\cite{qite_chan20, QITE_h2}.

As the current and near term NISQ hardware suffers short coherence time, gate infidelity, and other noises, the direct application of QITE on real devices is limited by the rather deep quantum circuits, in particular for systems with long-range correlations. The circuit depth grows linearly with the QITE steps, similar to the circuit to study the quantum dynamics following Trotter decomposition for the time-evolution operator~\cite{Trotter_dynamics_Lawrence, Trotter_dynamics_Knolle}. In contrast, the VQE calculations with an ansatz such as UCCSD features a variational circuit of fixed depth. In this paper, we develop a resource-efficient ``step-merged'' QITE (smQITE) algorithm, which performs approximate QITE calculations at fixed quantum circuit depth. The smQITE method builds on the numerical observations that the accumulated unitary operators in the QITE calculation can often be effectively combined. We will first present the smQITE formalism, followed by demonstrations that the smQITE method can produce high-quality results beyond chemical accuracy on a set of molecules. We demonstrate that the circuit depth of smQITE calculations can be further reduced significantly by adopting compact wavefunction representations, such as the UCCSD variational form among others, which effectively reduce the circuit depth down to that of UCCSD-VQE. It is shown that smQITE method can reach the accuracy of VQE with the same UCCSD ansatz, in much fewer steps without resorting to high-dimensional optimizations. Finally, we demonstrate the smQITE calculations for H$_2$ dimer on a real quantum device, with a binding energy curve in reasonable accuracy. We argue that, supported by numerical evidence, a combination of smQITE with VQE offers a way to address the highly complicated optimization problem of VQE when simulating large molecules.

\section{Step-merged QITE algorithm}
To be self-contained, we first review the quantum imaginary time evolution algorithm proposed by Motta et al~\cite{qite_chan20}, and point out the limitations for practical implementations on NISQ devices. The presentation of the step-merged QITE (smQITE) formalism then follows, which aims to dramatically reduce the circuit depth of QITE calculations on quantum computers, hence is better adapted for the current and near-term quantum devices.

\subsection{QITE algorithm}
 Consider an $N_\text{q}$-qubit system with Hamiltonian $\h=\sum_{m=0}^{M-1}{\hat{h}[m]}$, which includes a sum of $M$ weighted Pauli terms. The Pauli term $\hat{h}[m]$ is a general product of Pauli operators. The qubit Hamiltonian can naturally describe spin-$\frac{1}{2}$ models, or fermionic systems by mapping fermionic operators to qubit operators~\cite{map_bk, map_three}. Starting from an initial state $\ket{\Psi_0}$, the imaginary time evolution leads the system to the lowest eigenstate $\ket{\Psi_f}$ which has finite overlap with $\ket{\Psi_0}$ in the long time limit,

\bea
\ket{\Psi_f} &=& \lim_{\beta \rightarrow \infty}e^{-\beta \h}\ket{\Psi_0}.
\eea

The imaginary time evolution can be carried out through Trotter decomposition~\cite{trotter}
\bea
e^{-\beta \h} &=& (e^{-\Delta \tau \hat{h}[0]} e^{-\Delta \tau \hat{h}[1]} \cdots)^{N} \notag \\
&& +\O(\Delta \tau),   \label{trotter}
\eea
with the Trotter step size $\Delta \tau=\frac{\beta}{N}$.  Literally, the above evolution operator $e^{-\beta \h}$ consists of $M \times N$ steps, yielding an error of leading order proportional to $\Delta \tau$. For the convenience of discussions later, we label the Trotter step by $(n, m)\equiv n M + m$, with $0\leq n<N$ and $0\leq m<M$. The associated intermediate state is labelled as $\Psi_{(n,m)}$. After one additional Trotter evolution step, we have
\be
	\ket{\Psi_{(n,m)+1}} = c_{(n,m)}^{-\frac{1}{2}} e^{-\Delta\tau \hat{h}[m]}\ket{\Psi_{(n,m)}},
	\label{trotter-a}
\ee
The wavefunction norm is given by
\bea
	c_{(n,m)} &=& \Av{\Psi_{(n,m)}}{e^{-2\Delta\tau \hat{h}[m]}}  \notag \\
	&=& 1 -2\Delta\tau \Av{\Psi_{(n,m)}}{\hat{h}[m]} \notag \\
	&& + \O(\Delta \tau^2),
\eea
where to leading order in $\Delta \tau$ the deviation of the norm from unity is determined by the expectation value of the Hamiltonian in the intermediate state.

The main idea of QITE algorithm is to replace the non-unitary imaginary time Trotter evolution operator in Eq.~\eqref{trotter-a} by a unitary operator which transforms $\Psi_{(n,m)}$ to a state closest to $\Psi_{(n,m)+1}$,
\be
	\ket{\Psi_{(n,m)+1}} \approx e^{-i\Delta \tau \hat{A}^{(n,m)}} \ket{\Psi_{(n,m)}}.
\ee
Here, $\hat{A}^{(n,m)}$ is a Hermitian operator that can be expanded in a complete Pauli basis set of a domain of $D$ qubits around the support of $\hat{h}[m]$:
\be
	\hat{A}^{(n,m)} \equiv \sum_I {a^{(n,m)}_{I}\hat{\sigma}_I}.
	\label{amatrix}
\ee
Here, $I=i_0i_1...i_D$ is a composite index running through all the $D$ qubits. The domain $D$ includes at least all sites $m$, where $\hat{h}[m]$ acts non-trivially. Generally, the domain size $D$ can be larger than the support of a qubit operator due to correlation effects~\cite{qite_chan20}. The Pauli term $\hat{\sigma}_I=\hat{\sigma}_ {i_0}\hat{\sigma}_ {i_1}...\hat{\sigma}_ {i_D}$ is a product of Pauli operators. $\hat{\sigma}_i \in \{ I, X, Y, Z\}$ is a Pauli operator associated with the $i^{th}$ qubit. Without loss of generality, $a^{(n,m)}$ is a set of real parameters of dimension $4^D$ corresponding to rotation angles in the qubit Hilbert space. 

In order to determine the operator $\hat{A}^{(n,m)}$, we define the change of the state wavefunction after a Trotter imaginary time evolution step as
\bea
    \ket{\Delta^{(n,m)}_0} &=& \frac{\ket{\Psi_{(n,m)+1}}-\ket{\Psi_{(n,m)}}}{\Delta \tau}  \\
    &\approx& \left( \frac{c_{(n,m)}^{-\frac{1}{2}}-1}{\Delta \tau} - c_{(n,m)}^{-\frac{1}{2}}\hat{h}[m] \right)\ket{\Psi_{(n,m)}}, \notag
\eea 
where the Trotter exponential operator in Eq.\ref{trotter-a} is expanded to the first order of $\Delta \tau$. Similarly, for the unitary evolution we define the variation of the state as.
\bea
\ket{\Delta^{(n,m)}_1} &=& \frac{e^{-i\Delta \tau \hat{A}^{(n,m)}} \ket{\Psi_{(n.m)}}-\ket{\Psi_{(n,m)}}}{\Delta \tau} \notag \\
    &\approx& -i\hat{A}^{(n,m)}\ket{\Psi_{(n,m)}}.
\eea

The objective function to be minimized is defined as
\bea
f[a] &=& \olp{\Delta^{(n,m)}_0 - \Delta^{(n,m)}_1}  \\
    &=& f_0 + \sum_I {b_I a^{(n,m)}_I} + \sum_{I J}{a^{(n,m)}_I S_{I J} a^{(n,m)}_J} \notag
    \label{fa}
\eea
with 
\be
f_0 = \olp{\Delta^{(n,m)}_0},
\ee
\bea
    b_I &=& -i\Mel{\Delta^{(n,m)}_0}{\hat{\sigma}_I}{\Psi_{(n,m)}} + c.c, \\
    &\approx& i c_{(n,m)}^{-\frac{1}{2}}\Av{\Psi_{(n,m)}}{\hat{h}[m] \hat{\sigma}_I} + c.c., \notag
\eea
and
\be
S_{I J} = \Av{\Psi_{(n,m)}}{\hat{\sigma}_I^\dagger \hat{\sigma}_J}.
\ee
The minimization of the function $f[a]$ with respect to $a^{(n,m)}$ leads to a system of linear equations
\be
\left( \boldsymbol{S} + \boldsymbol{S}^T \right)\boldsymbol{a}^{(n,m)} = -\boldsymbol{b},
\label{le}
\ee
which is solved to determine the optimal expansion coefficients $a^{(n,m)}$ for the operator $\hat{A}^{(n,m)}$. Since $f_0$ does not enter the above linear equation, no explicit evaluation is needed. Quantum computers are employed to facilitate the setup of the linear equation (\ref{le}) by determining the $\boldsymbol{S}$-matrix and $\boldsymbol{b}$-vector. As the quantum computation only involves direct measurements of Pauli terms with respect to the state wavefunction, it is straightforwardly implemented on quantum devices. The number of linear equations in Eq.~\eqref{le} is $4^D$, which scales exponentially with the number of qubits $D$ in the relevant qubit domain. With increasing system size, this rapidly becomes the bottleneck of the algorithm. We will discuss alternative ways to lift this constraint in section~\ref{smQITE_VQE}.

\subsection{Step-merged QITE}
\begin{figure}[ht!]
	\centering
	\includegraphics[width=0.45\textwidth]{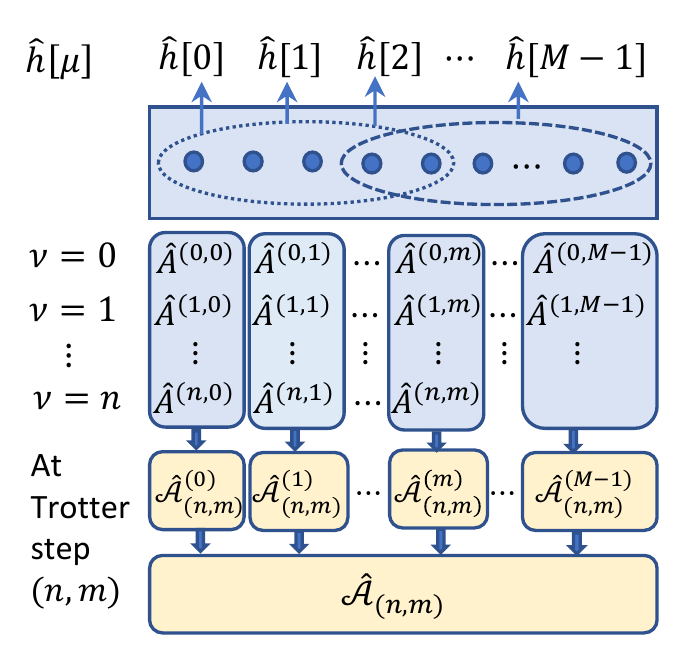}
	\caption{
	\textbf{Schematic illustration of combining Trotter unitaries in smQITE algorithm.} At Trotter step $(n,m)$ with imaginary time evolution operator $e^{-\Delta \tau \hat{h}[m]}$, a new unitary operator $e^{-i\Delta \tau \hat{A}^{(n,m)}}$ is appended to the quantum circuit. The operator $\hat{A}^{(n,m)}$ is defined in a qubit domain $D_m$ around the support of $\hat{h}[m]$. A set of Pauli terms in the qubit Hamiltonian can share a common qubit domain, as indicated by the dotted ellipse. As the accumulated operators $\{ \hat{A}^{(\nu,\mu)} \}$ at Trotter step $(n,m)$ share the same qubit domain if they have the same column index $\mu$, they can be combined to $\hat{\mathcal{A}}^{(\mu)}_{(n,m)} = \sum_\nu A^{(\nu, \mu)}$. By defining a union of the Pauli basis set in all qubit domains $D = D_0 \cup \ldots \cup D_M$, the operators $\{\hat{\mathcal{A}}^{(\mu)}_{(n,m)} \}$ can be further combined to a single operator $\hat{\mathcal{A}}_{(n,m)} = \sum_{\mu = 0}^{M-1} \hat{\mathcal{A}}^{(\mu)}_{(n,m)}$.
	}
	\label{figure0}
\end{figure}

A key factor in determining the required quantum resources of the QITE approach is the preparation of state $\Psi_{(n,m)}$ at Trotter step $(n,m)$, which will be repeated for all the measurements. The state $\Psi_{(n,m)}$ is constructed as
\bea
\ket{\Psi_{(n,m)}} &=& \prod_{\mu'=0}^{m}e^{-i\Delta \tau \hat{A}^{(n,\mu')}} \notag \\
  &&\times\prod_{\nu=0}^{n-1}\prod_{\mu=0}^{M -1}e^{-i\Delta \tau \hat{A}^{(\nu,\mu)}}\ket{\Psi_0}, \label{ltrotter}
\eea
where the exponential operators are ordered according to the Trotter evolution path, as also illustrated in Fig.~\ref{figure0}. Clearly, the depth of the state preparation circuit grows linearly with the Trotter steps, which limits the system size and maximal Trotter steps that the QITE algorithm can perform in NISQ devices. In contrast, the variational quantum algorithms, such as variational quantum eigensolver with unitary coupled cluster ansatz~\cite{vqe_theory, vqe_uccsd}, have an advantage of a variational quantum circuit at fixed depth. Although some approximate ways have been discussed in references~\cite{qite_chan20, QITE_h2, qite_nla}, the linear growth of the quantum circuit depth with increasing Trotter steps has not been addressed. 

Here, we propose a step-merged QITE (smQITE) approach to control the circuit depth at an effective single (or few) Trotter step level. The key idea is to combine Trotter evolution unitaries along the state evolution path, which act on a common set of qubits. The algorithm is schematically depicted in Fig.~\ref{figure0}. While this heuristic approach does not become exact in the limit $N\rightarrow\infty$, we show below that it leads to results for ground state energies that are comparable to VQE. This is remarkable as, unlike VQE, the smQITE approach does not require performing a difficult optimization in a high-dimensional feature space. We further discuss a systematic way to improve the accuracy of smQITE at the cost of using deeper circuits. Finally, the smQITE method can also be combined with VQE, as it yields an efficient ansatz for the ground state that can be further optimized variationally.

More specifically, by commuting terms with a common index $\mu$  next to each other in Eq.~\eqref{ltrotter}, we can rewrite the state evolution in this equation as
\be
\ket{\Psi_{(n,m)}} = e^{-i\Delta \tau \sum_{\mu=0}^{M -1} \hat{\mathcal{A}}^{(\mu)}_{(n,m)}}\ket{\Psi_0} + \mathcal{O}(\Delta \tau^2) \,. \label{strotter}
\ee
Here, we have defined  $\hat{\mathcal{A}}^{(\mu)}_{(n,m)} = \sum_{\nu=0}^{n} A^{(\nu, \mu)}$ for $\mu \leq m$. For $\mu > m$, the summation stops at $\nu=n-1$. This expression combines the operators $\hat{A}^{(\nu,\mu)}$ with a common index $\mu$ that share the same Pauli basis in the qubit domain $D_\mu$ around the support of $h[\mu]$. By commuting the exponential terms to bring terms with a common $\mu$ next to each other, we have generated a number of terms that are all of the order of $\Delta \tau^2$. We discuss the issue of the Trotter error in more detail below.

Further grouping is possible if different qubit domains $D_\mu$ of different $\hat{h}[\mu]$ overlap and some $\hat{\mathcal{A}}^{(\mu)}$-operators can be further combined. Without loss of generality, we define an extended Pauli basis set $\{\sigma_I\}$ as the union of all the Pauli basis sets in the different qubit domains $D_\mu$ of the Hamiltonian terms $\{ \hat{h}[\mu] \}$. This allows us to maximally combine the operators $\hat{\mathcal{A}}_{(n,m)} \equiv \sum_{\mu=0}^{M -1} \hat{\mathcal{A}}^{(\mu)}_{(n,m)}$ and represent it in the extended Pauli basis set, as illustrated in Fig.~\ref{figure0}. The smQITE wavefunction at Trotter step $(n,m)$ is then given by 
\begin{equation}
    \ket{\Psi_{(n,m)}} = e^{- i \Delta \tau \hat{\mathcal{A}}_{(n,m)}} \ket{\Psi_0}\,,
    \label{eq:Psi_smQITE_all_combined}
\end{equation}
which corresponds to a single effective Trotter step. Note that in the original QITE paper~\cite{qite_chan20}, $\hat{h}[m]$ is defined according to operational locality and can be a sum of Pauli terms sharing a common qubit domain. Therefore, an effective combination of $\hat{A}^{(\nu,\mu)}$ over index $\mu$ at a common qubit domain has been performed, albeit at each individual Trotter step $\nu$.

In the case of \textit{ab initio} molecular Hamiltonians where long-range one-body and two-body operators are present,
it is often the case that the set of $\{\hat{h}[\mu]\}$ Hamiltonian terms share a common domain of qubits, that often spans the full system. It is thus natural to consider the evolution under the full $\hat{\h}$: 
\be
	\ket{\Psi_{(n+1)}} = c_{(n)}^{-\frac{1}{2}} e^{-\Delta\tau \h}\ket{\Psi_{(n)}},
	\label{trotter-b}
\ee
rather than Eq.~\eqref{trotter-a}. Note that we do not introduce the domain index $\mu$ as there is only a single domain spanning the full system. The state evolution  Eq.~\eqref{strotter} thus reads as
\be
\ket{\Psi_{(n)}} = e^{-i\Delta \tau \hat{\A}_{(n)}}\ket{\Psi_0}, \label{wfn}
\ee
with $\hat{\A}_{(n)} = \sum_{\nu = 0}^{n-1} \hat{A}^{(\nu)}=\sum_I\left(\sum_{\nu=0}^{n-1} a_I^{(\nu)}\right)\hat{\sigma}_I$, where we have combined Trotter unitaries with different step index $\nu$. The step of combined Trotter evolution of the state wavefunction across the whole set of $\{\hat{h}[\mu] \}$ does not change the quantum circuit depth. However, it potentially saves time for systems with largely overlapping qubit domains $\{D_\mu\}$, such as molecules, due to the prevalence of nonlocal one-body and two-body operators. Furthermore, it introduce a new perspective that a compact representation of the $\hat{\A_{n}}$ operator can be obtained through variational wavefunction forms of VQE, which will be detailed in section~\ref{smQITE_VQE}. It has been discussed  recently that the Pauli operator ordering in the Trotterized circuits of the VQE-UCCSD approach can introduce significant errors in energy evaluations beyond chemical accuracy~\cite{VQE_order}. As a Trotterized form is also adopted in the smQITE method, similar operator ordering effects could exist. Nevertheless, we will demonstrate that decent numerical results from smQITE calculations can already be obtained without exploiting optimum Pauli operator orderings. For the purpose of reproducibility of numerical results, all our calculations, including explicit ordered list of Pauli operators, are publicly accessible in the online repository~\cite{smQITEData}.

As the number of Trotter steps $N$ increases, the smQITE approach maintains a favorable fixed circuit depth. This is in stark contrast to the linear growth of the depth with $N$ found in QITE~\cite{qite_chan20}. But the gain in quantum resource efficiency is obtained at a price. In the worst case scenario where none of the operators $\{\hat{A}^{(\nu,\mu)}\}$ commute with each other and all leading Trotter errors are of the same sign and add up, the above step merging procedure introduces a constant error. The smQITE approach thus loses the mathematical rigor of QITE and does not become exact in the limit of small Trotter step size $\Delta \tau = \beta/N \xrightarrow{N \rightarrow \infty} 0$. The smQITE method should thus be regarded as a heuristic approach that can still work well in the average case, as we demonstrate for a number of examples below. Even in this worst case scenario where the Trotter error is uncontrolled, the energy obtained from the smQITE ansatz is still a variational upper bound, and the smQITE wavefunction $\ket{\Psi_{(n,m)}}$ in Eq.~\eqref{eq:Psi_smQITE_all_combined} can be used as a starting point for further variational optimization using VQE. It is worth noting that the operator $\hat{A}^{(n,m)}$ in Eq.~\eqref{amatrix} is first determined variationally at each smQITE step, and subsequently merged into the preceding unitary operators. In other words, the QITE procedure is followed initially, but in order to avoid a further growth of the circuit depth the preparation of state $\ket{\Psi_{(n,m)}}$ is approximately achieved by using the step-merged unitary in Eq.~\eqref{eq:Psi_smQITE_all_combined}. Therefore, the effective single-step smQITE ansatz is generally different form the a QITE ansatz with a single Trotter step. In fact, because the smQITE approach coincides with QITE at the first Trotter step where no combination of unitaries has been performed, smQITE can always achieve the single-step QITE result as an upper bound. The error in smQITE calculations should be equal or smaller than that by effectively reducing the Trotter decomposition in Eq.~\eqref{trotter} from order $N$ to order $1$, which will also be demonstrated numerically in section~\ref{smQITE_CPB}.

Finally, let us describe a way to detect the Trotter errors induced by the step merging process and a way to iteratively reduce it. One way to estimate this error is to compare the energy of the state obtained from merging all Trotter steps $\nu$ into a single effective step, $\hat{\mathcal{A}}^{(\mu)}_{(n,m)} = \sum_{\nu = 0}^n A^{(\nu, \mu)}$, versus merging them into two effective steps, $\hat{\mathcal{A}}^{(\nu, \mu)}_{(n,m)} = \sum_{\nu' = \frac{n}{2} \nu}^{\frac{n}{2}(\nu + 1)} A^{(\nu', \mu)}$ with $\nu = 0, 1$. If the energy decreases when using more effective Trotter steps, this process can be repeated until convergence. Obviously, this process approaches the original QITE limit if we increase the range of the index $\nu$ and hence requires increasingly deep circuits to prepare the wavefunction $\ket{\Psi_{(n,m)}}$.

\section{Application of step-merged QITE to quantum chemistry}
In this section, we show that highly accurate results beyond chemical accuracy can be obtained for the smQITE calculations for a set of molecules. In particular, we prove numerically that the high accuracy of smQITE method cannot be obtained by instead using a single Trotter step calculation, even when using an optimum step size $\Delta \tau$. We further propose a way to effectively adopt the variational wavefunction form of VQE into smQITE. For a number of molecules, we show that smQITE yields results of similar accuracy as VQE with the same fixed variational ansatz, yet with much fewer steps and shallower circuits. Finally, we report results of smQITE calculations performed on Rigetti QPUs.

\subsection{Implementation of smQITE for quantum chemistry}
Consider an \textit{ab initio} nonrelativistic molecular electron Hamiltonian
\bea
\h &=& \sum_{p q}\sum_{\sigma}h_{p q}\cc_{p\sigma}\ca_{q\sigma} \notag \\
   &+& \frac{1}{2}\sum_{p q r s}\sum_{\sigma \sigma'}h_{p q r s}\cc_{p \sigma} \cc_{r \sigma'} \ca_{s \sigma'} \ca_{q \sigma}, \label{mh}
\eea
with the one-electron core part of the Hamiltonian given by
\be
h_{p q} = \int d \br \phi_{p}^{*}(\br)(\T + \V_{ion})\phi_q(\br),
\ee
and the two-electron Coulomb integral 
\be
h_{p q r s} = \int d \br \int d \br' \phi_{p}^{*}(\br)\phi_{r}^{*}(\br')\V_{e e}\phi_s(\br')\phi_q(\br).
\ee
Here $p,q,r,s$ are composite indices for atom and orbital, and $\sigma$ is spin index with values of $\alpha$ for spin-up and $\beta$ for spin-down. $\T$ is the kinetic energy operator, $\V_{ion}$ is the ionic potential operator and $\V_{e e}$ the Coulomb interaction operator. $\{\phi(\br)\}$ is a set of basis orbital functions, which are obtained from the standard STO-3G minimal basis set. In the following smQITE calculations of molecules, a quantum chemistry package \textit{PySCF} is first used to get the restricted Hartree-Fock(HF) solution~\cite{pyscf_sun2018}. The molecular Hamiltonian (Eq.~\ref{mh}) is then transformed to the molecular orbital representation for the convenience of preparation of the initial HF state in quantum computer. The qubit representation of the Hamiltonian is obtained by parity transformation, with two qubits reduced by exploiting the conservation of total number of electrons and $Z$-component of the total spin operator, e.g., the $Z_2$ symmetry. The smQITE code is implemented using modules from Qiskit~\cite{Qiskit} and Forest~\cite{pyquil_0, pyquil_1}, and is available as a module in the open-source package PyGQCE~\cite{pygqce}. The smQITE method is a general Hamiltonian eigensolver, with potential applications beyond quantum chemistry problems, such as the impurity models~\cite{gqce}.


\subsection{smQITE calculations using a complete Pauli basis set}
\label{smQITE_CPB}

\begin{figure*}[th!]
	\centering
	\includegraphics[width=\linewidth]{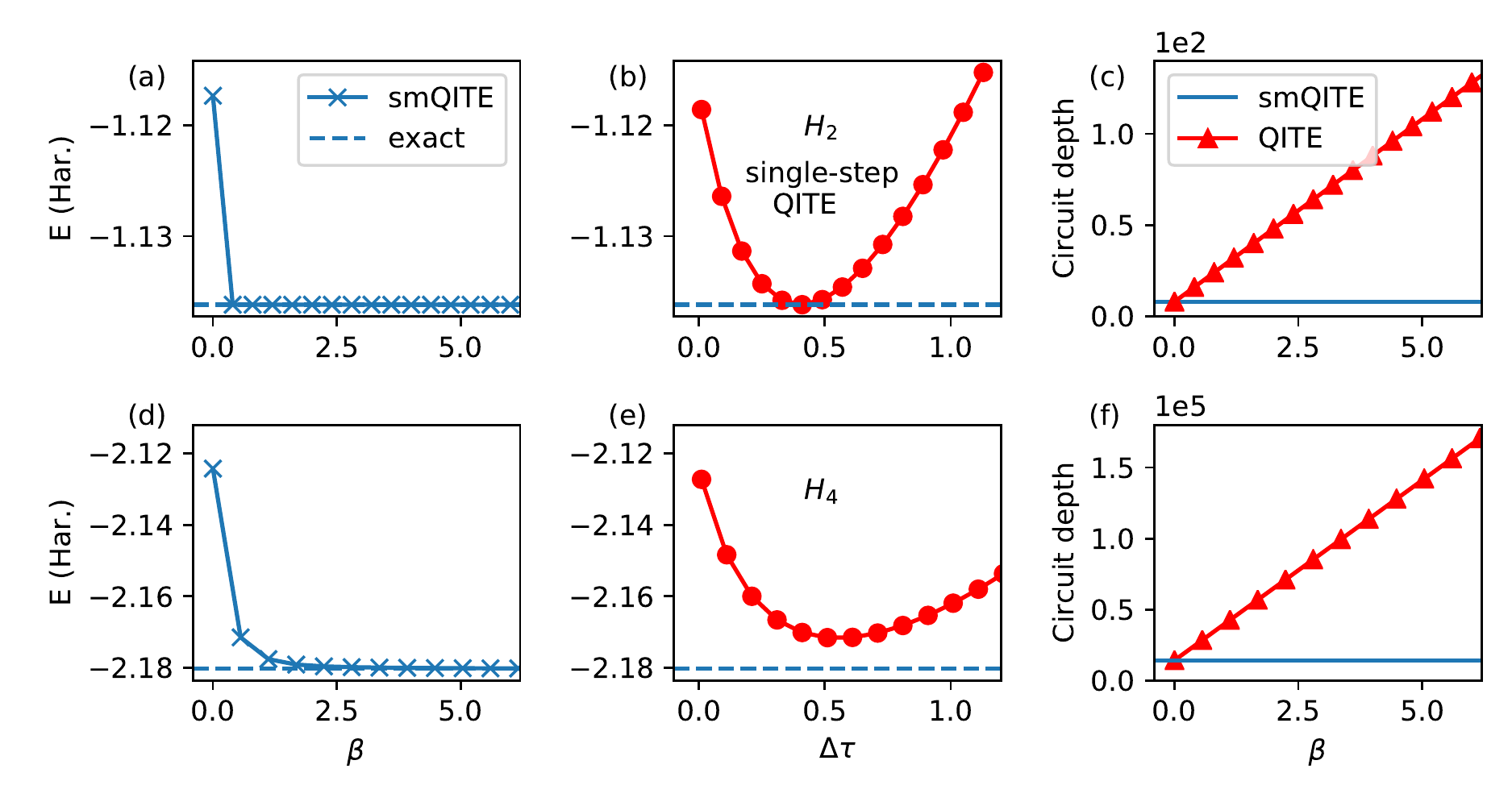}
	\caption{
	\textbf{Energy convergence and fixed circuit depth of the smQITE method.} Upper panels show the energy evolution as a function of merged QITE steps (a), the energy as a function of single QITE step size (b), and the quantum circuit depth of QITE and smQITE calculations (c) for H$_2$ dimer at bond length $0.7$\AA, together with the results for H$_4$ chain at bond length $0.9$\AA\,in lower panels. Note that the circuit depth at order of $10^5$ is far beyond the capability of the current NISQ devices.
	}
	\label{figure1}
\end{figure*}

Figure~\ref{figure1} shows the evolution of the Hamiltonian expectation value $E$ as a function of $\beta=n\times \Delta \tau$ for H$_2$ dimer and H$_4$ chain in panel (a) and (d), which quickly converges to the exact result from the initial value of the HF solution. In the middle panels (b) and (e), we plot the energy $E$ after a single QITE step upon the initial HF wavefunction with varying the Trotter step size $\Delta \tau$, which shows a polynomial behavior with a unique minimum $E_{min}^{(1)}$ at an optimal step size $\Delta \tau_{opt}$. Accidentally, $E_{min}^{(1)}$ coincides with the exact energy for H$_2$, which is due to the simple structure of the Hamiltonian. Generally, $E_{min}^{(1)}$ will be higher than the exact result. For the case of H$_4$, the energy is overestimated by 5 $kcal/mol$, which is beyond the chemical accuracy of 1 $kcal/mol$~\cite{pople1999nobel}. For comparison, $\Delta \tau_{opt}$ is used as the fixed step size $\Delta \tau$ for the smQITE calculations. Fig.~\ref{figure1}(b) clearly shows that the smQITE calculation of H$_4$ can reach a much higher accuracy (~$10^{-4}kcal/mol$) after a few steps. The calculations are performed on a wavefunction simulator as implemented in Forest~\cite{pyquil_0, pyquil_1}, which is equivalent to perfect measurements on fault-tolerant quantum computers. We estimate the quantum circuit depth by counting the number of two-qubit controlled-NOT (CNOT) gates in the algorithms, which are shown in Fig.~\ref{figure1}(c) and (f) for calculations of  H$_2$ and H$_4$, respectively. As expected, the smQITE circuit has a fixed depth at 8 for H$_2$ and 14208 for H$_4$. In contrast, the QITE circuit grows linearly in depth as the QITE step proceeds.

\begin{figure*}[th!]
	\centering
	\includegraphics[width=\linewidth]{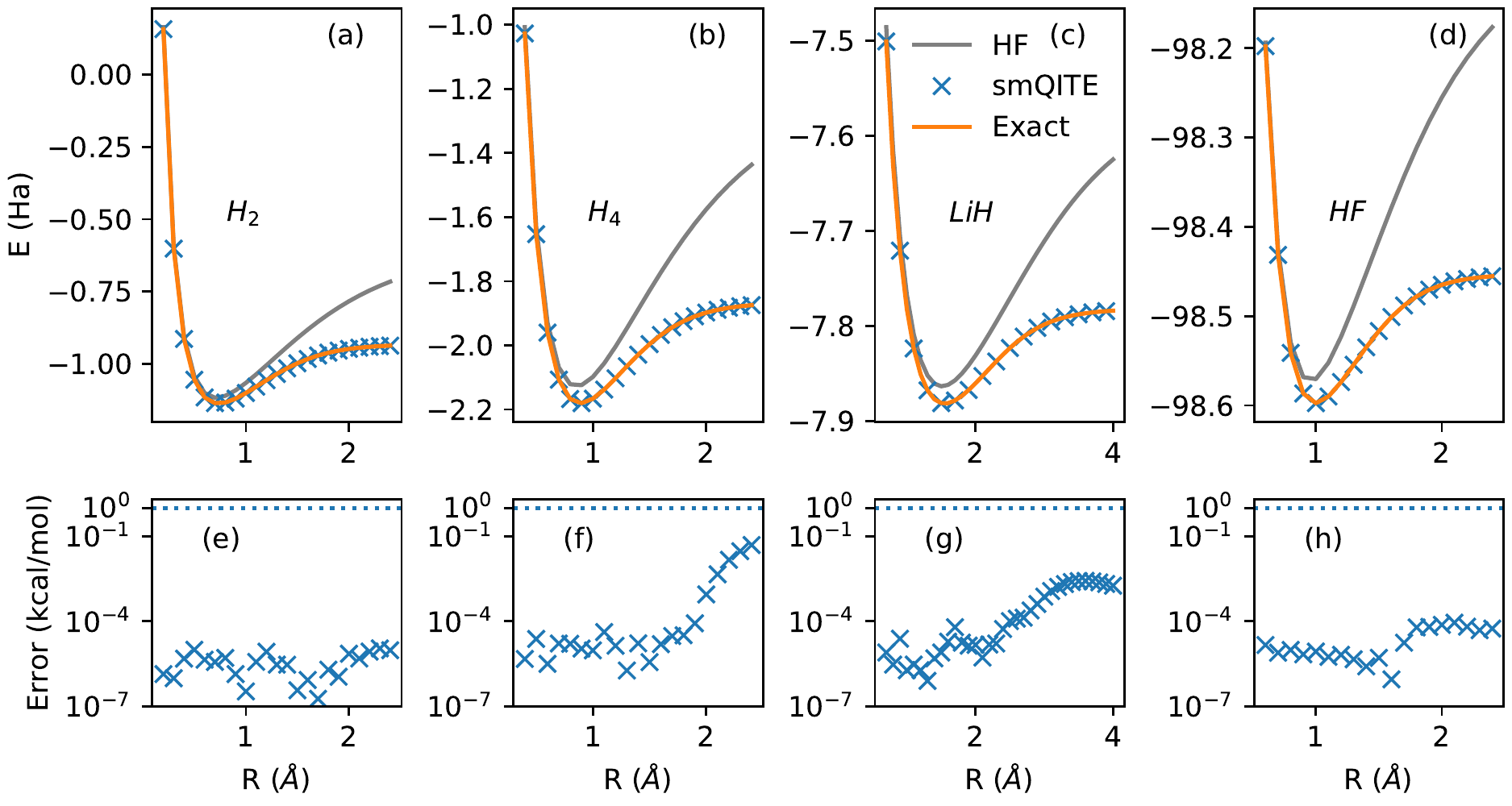}
	\caption{
	\textbf{Binding energy curves from smQITE calculations.} The binding energy curves of H$_2$ (a), H$_4$ chain (b), LiH (c) and HF (d) molecules from smQITE calculations are plotted together with exact and HF results. The error of smQITE calculations, $E_{smQITE} - E_{Exact}$, as shown in panels (e-h), are well below the chemical accuracy threshold, as indicated by the horizontal dotted line in the lower panels. 
	}
	\label{figure2}
\end{figure*}

We further apply the smQITE method to a set of molecules to map out the full binding and dissociation energy curves, which give a more complete assessment of the computational accuracy. The smQITE results are reported with the exact curves in Fig.~\ref{figure2} for molecules H$_2$(a), H$_4$(b), LiH(c) and HF(d). The associated error, defined as the energy difference between the smQITE and exact diagonalization (ED, or full configuration interaction, FCI) calculations, is plotted in the lower panels (e-h). In all the cases, the smQITE calculations yield energies in much better agreement with the exact answers beyond the chemical accuracy. The Hartree-Fock binding energy curves have also been shown for reference, which provides a measure for the electron correlation effects in the system. For polyatomic molecules composed of atoms with open-shell, such as H, Li and F atom, the correlation energy, defined as the energy difference between Hartree-Fock and exact calculations, increases as the molecule is uniformly stretched toward the dissociation limit. The smQITE method recovers almost all the correlation energy. 

In the Hartree-Fock calculations for LiH molecule, the STO-3G minimal basis set describes $1s$-orbital for H and $1s$, $2s$, and $2p$-orbitals for Li. The Li $1s$ orbital is kept in the core, as it is fully occupied and deep in energy level. The $2p_y$ and $2p_z$-orbitals are discarded because they do not participate in bonding and remain empty due to the symmetry constraints for the geometry aligned along $x$-axis. Therefore, four qubits are needed to represent the LiH Hamiltonian with $Z_2$ symmetry. In the case of HF molecule, the minimal basis contains H $1s$-orbital and F $1s$, $2s$, and $2p$-orbitals. Here we keep all the orbitals in the calculations, except F $1s$ and $2s$-orbitals, as they are much deeper in the core. Thus six qubits are used to represent the Hamiltonian of HF molecule, like the simulation of H$_4$. The detailed setup of the calculations can be found in online repository~\cite{smQITEData}.

\subsection{smQITE calculations using a compact Pauli basis set}
\label{smQITE_VQE}

\begin{figure*}[th!]
	\centering
	\includegraphics[width=\linewidth]{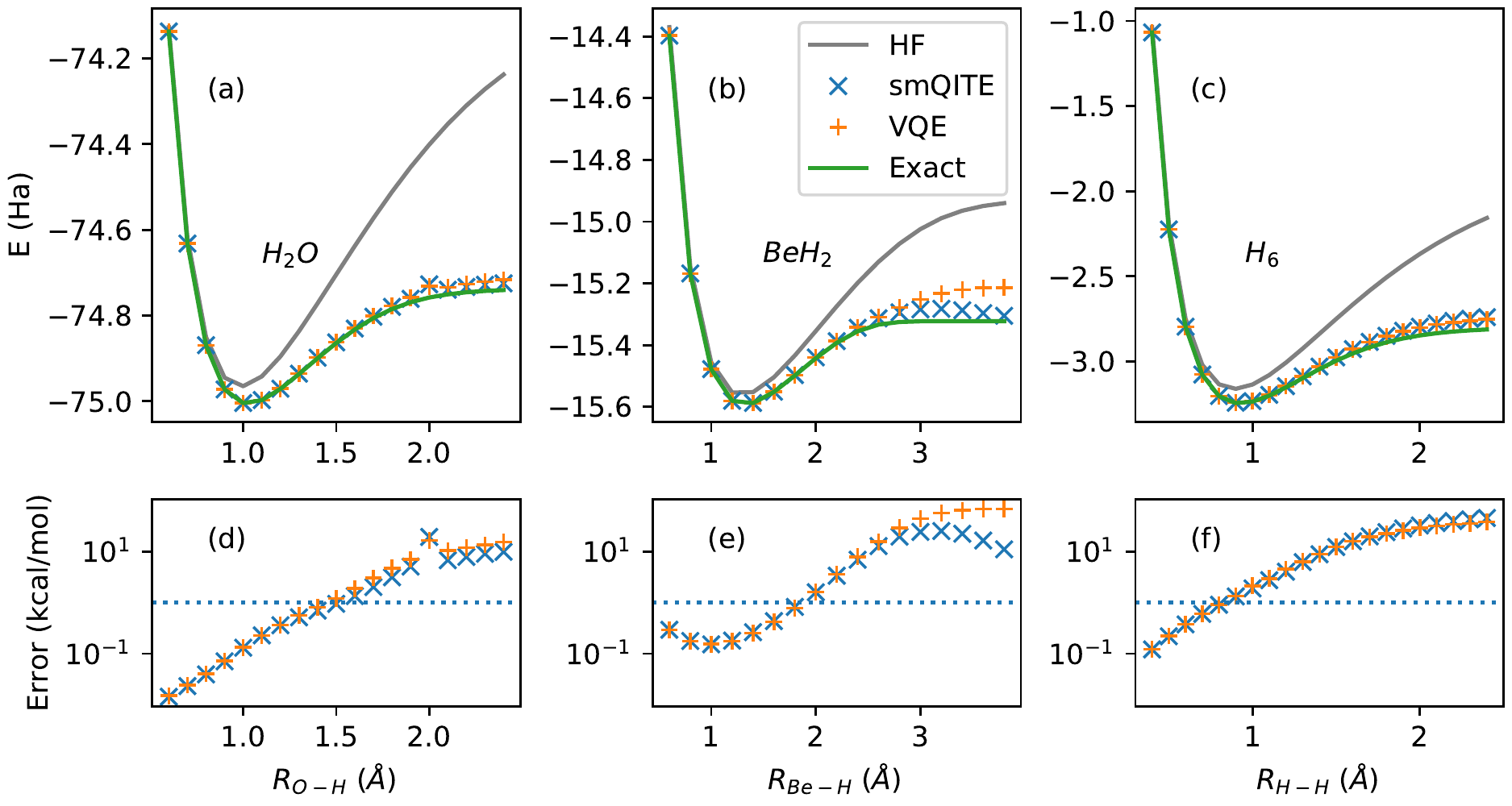}
	\caption{
	\textbf{Application of smQITE to molecules using simplified UCCSD ansatz.} The binding energy curve from smQITE calculations is shown for H$_2$O in panel (a), BeH$_2$ in (b) and H$_6$ chain in (c). The results from exact diagonalization, VQE with the same simplified UCCSD ansatz, and HF calculations are also shown for comparison. The errors of smQITE and VQE calculations, which measure the difference from the exact answers, are plotted in panels (d-f).
	}
	\label{figure3}
\end{figure*}

A limitation in the above smQITE calculations is that the dimension of the system of linear equations~\eqref{le} grows exponentially as $4^D$ with respect to qubit domain size $D$ determined by the electron correlations. To simulate systems of increasing size, some approximate treatment has been introduced in the reference.$^{25}$ Specifically, QITE calculations can be performed with a reduced qubit domain size $D'$, by choosing a subset of Pauli terms of length $L\leq D'$ to represent $\hat{A}$ in Eq.~\eqref{amatrix}. This approximation becomes equivalent to mean-field solution for $D'=1$ and approaches to exact result with increasing $D'$. Approximate QITE calculations have been demonstrated to be quite effective for 1D short-range spin models up to 20 qubits, as well as for 1D long-range Heisenberg Hamiltonian, albeit of much shorter 6 qubits.

The \textit{ab initio} molecular Hamiltonian usually has a much more complex structure than the spin models aforementioned, due to the presence of long-range one-body hopping and two-body interaction terms. Hence the qubit domain associated with a Pauli term in the Hamiltonian could be significantly larger. For example, the qubit representation of the electron Hamiltonian of H$_4$ molecule contains a Pauli term which acts on all the qubits, independent of the choice for encoding: Jordan-Wigner, parity or Bravyi-Kitaev transformation~\cite{map_bk, map_three}. As a result, the qubit domain should include all the qubits in the calculations, as adopted in the smQITE calculations reported before. Note that the operator domain size is dependent of specific representations. For fermionic systems, the linear equation~\eqref{le} can also be constructed using fermionic representation of $\hat{h}[m]$ and $\hat{A}$, which will be expanded in the basis of tensor product of fermionic operators $\{I, \ca_{p\s}, \cc_{p\s}, \cc_{p\s}\ca_{p\s} \}$~\cite{qite_chan20}. In this fermionic representation, the domain size of $\hat{h}[m]$ generally depends on the many-body state it acts upon, and the length of Pauli term of $\hat{h}[m]$ in qubit representation becomes irrelevant.

To extend the application of smQITE to molecules of increasing size, we propose an alternative approach to reduce the computational complexity. The dimension of the system of linear equations~\eqref{le} can be effectively reduced by choosing an optimal subset of Pauli basis for the representation of Hermitian operator $\hat{A}$ in Eq.~\ref{amatrix}. Note that the smQITE approach produces a wavefunction ansatz in Eq.~\ref{wfn}, which resembles the variational wavefunction form of VQE, such as the UCCSD ansatz in qubit representation
\bea
\ket{\Psi(\Vec{\theta})} &=& e^{\hat{T}(\Vec{\theta})-\hat{T}^\dagger(\Vec{\theta})}\ket{\Psi_0} \notag \\ 
&=& e^{-i\sum_j \theta_j f_j(\{\hat{\sigma} \}) }\ket{\Psi_0}. \label{ucc}
\eea
Here $f_j(\{\hat{\sigma} \})$ is a weighted sum of Pauli terms associated with the $j^{th}$ fermionic operator for the single or double excitation. However, the UCCSD ansatz includes much fewer Pauli terms, which naturally provides an alternative compact Pauli basis set, rather than a complete Pauli basis set of exponentially growing dimension $4^D$, for the representation of the Hermitian $\hat{A}$ operator in Eq.~\ref{wfn}, and equivalently reduces the dimension of the system of linear equations~\eqref{le}. 

As each $f_j(\{\hat{\sigma} \})$ in Eq.~\ref{ucc} usually includes several Pauli terms (2 for single excitations and 8 for double excitations), this translates to a quite significant overhead for the quantum circuit. Indeed, it has been demonstrated that reformulating the exponential ansatz~\eqref{ucc} utilizing directly the qubit evolution operators (Pauli terms) leads to a generally much shallower circuit~\cite{VQE_qcc, MayhallQubitAVQE}. However, the introduced overhead for simulations is a screening process for selecting qubit operators, which inevitably renders the ansatz system-dependent and lose the generality of the wavefunction form of the UCCSD ansatz in Eq.~\ref{ucc}. Here we take an alternative approach to simplify UCCSD ansatz preserving the general wavefunction form without operator-screening. The proposal is to replace $f_j(\{\hat{\sigma} \})$ by one of the list of Pauli terms in $f_j$~\cite{FengVQE}. The advantage is that it preserves the general variational wavefunction form and extremely easy to implement based on an existing UCCSD code. Although the simplified UCCSD (sUCCSD) ansatz remains generally subject to static correlation error as the UCCSD ansatz, it serves well our purpose here to demonstrate that adopting the compact list of Pauli operators in the UCC-type exponential ansatz enables quite accurate smQITE calculations of molecules with increasing size. In numerical examples to be discussed below, we do not find a significant effect on the specific choice of Pauli term in $f_j(\{\hat{\sigma} \})$ based on our preliminary tests. A systematic study on the optimum choice of Pauli terms and the effect on the quantum circuit structure and numerical accuracy is of interest and will be addressed in future work. The details of our calculations can be found in the open repository~\cite{smQITEData}. We include explicitly lists of Pauli basis set ordered according to real calculations for reference, since it has been demonstrated recently that different qubit operator orders in VQE calculations with the Trotterized form of UCC ansatz could affect final results quite significantly~\cite{VQE_order}. We note that the variational ansatz-based quantum simulation of imaginary time evolution (VQITE) recently proposed by McArdle, et al resembles our smQITE method with representations from VQE ansatz in some aspect~\cite{VQITE}. However, VQITE is derived using McLachlan’s time-dependent variational principle and the guiding equations are completely different~\cite{variational_mclachlan, variational_equivalence}. Furthermore, the evaluation of coefficients in the VQITE equation of motion on quantum computers introduces additional overhead of an ancillary qubit and generally complicated controlled-unitary operators~\cite{VQITE, VDynamics_Li}.

We demonstrate the smQITE calculations with the above sUCCSD Pauli operator set on molecules H$_2$O, BeH$_2$ and H$_6$, as shown in Fig.~\ref{figure3}. The binding energy curves from exact diagonalization and VQE calculations with the same sUCCSD ansatz are also shown for comparison. The HF results are given as a reference to estimate the dynamic and static correlation effects. The smQITE calculation results generally stay in close agreement with VQE calculations, and they both reach chemical accuracy when the bond length near or smaller then the energetically optimum value, where the dynamical correlation effect dominates. As the bond length increases towards the dissociation limit where the static correlation takes over, the errors start to go beyond chemical accuracy, due to the single reference nature of the sUCCSD ansatz. The smQITE and VQE binding curves are generally very smooth, except one energy point of H$_2$O at O-H bond length of 2.0\AA, which we attribute to a possible limitation of the sUCCSD variational wavefunction form. We expect that more sophisticated variational forms, such as $k$-UpCCGSD~\cite{kUpUCCGSD} or UCC with paired double excitations plus orbital optimization~\cite{oopUCCD, qoopUCCD}, may give better compact Pauli representation for smQITE calculations, and improve the accuracy near dissociation limit.

In QITE or smQITE calculations, the Trotter step size $\Delta \tau$ can significantly affect the convergence speed of the Hamiltonian expectation value. Generally, $\Delta\tau$ can be gradually increased for molecules with increasing bond length for faster convergence, where static correlation effects become stronger. Take the smQITE calculation of H$_4$ in Fig.~\ref{figure2} as an example. It takes only 3 smQITE steps to reach chemical accuracy with $\Delta\tau=0.2$ for H$_4$ at bond length $R=0.7$\AA, while it takes 22 steps to converge to chemical accuracy with the same step size at $R=2.4$\AA. If we choose a bigger $\Delta\tau=1.5$, it takes only 4 steps to reach the chemical accuracy. Although the optimum $\Delta\tau$ is system-dependent and not known a priori, smQITE calculations with auto-tuned $\Delta\tau$ can be easily implemented. More precisely, it is feasible to choose a large enough initial value for $\Delta\tau$ to start the smQITE calculation. The energy at each smQITE step is monitored. If the energy starts to increase, $\Delta\tau$ will be scaled down by a constant factor (e.g., 5) and the smQITE solution returns to the lowest energy point achieved in the previous steps. The smQITE calculation then continues with the updated $\Delta\tau$, which can be further reduced accordingly. The smQITE calculation terminates if $\Delta\tau$ is sufficiently small (e.g., $\Delta\tau < 1.e^{-4}$) or energy converges to the desired accuracy. The smQITE calculations for H$_2$O, BeH$_2$ and H$_6$ in Fig.~\ref{figure3} are carried out with the Trotter step size $\Delta\tau$ dynamically adjusted as described above. All the calculations converge in energy of $0.1m Ha$ within 80 steps. In contrast, the VQE calculations require from several hundred up to two thousand steps to achieve similar convergence, if the sequential least squares programming (SLSQP) optimization method is used. Significantly more steps are necessary if the sUCCSD ansatz is optimized using constrained optimization by linear approximation (COBYLA) method. 

Rigorously speaking, the VQE step, characterized by the calculation of Hamiltonian expectation value with respect to an updated wave function, can take much less time than the smQITE step, as many additional terms defining Eq.~\ref{le} must be evaluated in the smQITE method. Consequently, the computational time of smQITE and VQE calculations is comparable. For example, it takes about 102 seconds for smQITE and 188 seconds for VQE calculation of the H$_6$ chain at $R=1.4$\AA\,with an Intel Xeon Processor(Skylake, IBRS). However, all the measurements at each step can potentially be performed in parallel as they are independent. Moreover, the optimization of the variational ansatz is generally a non-convex problem, and can be very challenging to reach the global minimum within a high-dimensional parameter space given by many variational parameters. In contrast, the smQITE calculation proceeds along a well-defined imaginary time evolution path, which is free of the potential complications of high-dimensional non-convex optimization problems. As shown in Fig.~\ref{figure3}(b), the smQITE calculation gives appreciably lower energy than VQE for BeH$_2$ close to dissociation limit. In principle, VQE should always lead to an energy, which is the same or lower than the smQITE result at the global minimum in its variational space, given that both approaches share the same variational wavefunction form. In fact, VQE can further improve the smQITE energy if the smQITE solution is used as the starting point for the variational optimization. For example, the final energy can be further improved by more than 2 mHa for BeH$_2$ at bond length of 3.8\AA. This suggests that a combination of smQITE and VQE may offer a way to overcome the challenge of high-dimensional non-convex optimization problem inherent in the VQE approach. Note that the convergence of VQE calculations can also be improved by utilizing the analytical gradient of the cost function. However, the evaluation of gradient on quantum computers introduces the similar overhead of an ancillary qubit and controlled-unitary operators as in the VQITE method mentioned above~\cite{alan_ucc2018, VQITE}.

In the above Hartree-Fock calculations for H$_2$O molecule, the STO-3G minimal basis set describes $1s$-orbital for H and $1s$, $2s$, and $2p$-orbitals for O. The O $1s$ and $2s$ orbitals are kept in the core, as they are fully occupied and deep in energy level. Therefore, eight qubits are needed to represent the H$_2$O Hamiltonian with $Z_2$ symmetry. In the case of BeH$_2$ molecule, the minimal basis contains H $1s$-orbital and Be $1s$, $2s$, and $2p$-orbitals. Here we keep Be $1s$ orbital in the core and remove Be $2p_z$ as it doesn't participate in bonding for the molecule aligned in xy-plane. Therefore, eight qubits are used to represent the Hamiltonian of BeH$_2$ molecule, like the simulation of H$_6$. The number of Pauli terms in the Hamiltonian of H$_2$O, BeH$_2$ and H$_6$ is 252, 252, and 919, and the associated number of variational parameters of the sUCCSD ansatz, or equivalently the dimension of the Pauli basis in smQITE calculations, is 54, 54, and 59, respectively. Remarkably, the smQITE calculation with sUCCSD ansatz for molecules performs much better than the previously proposed sparse representation based on reducing the qubit domain size of $\hat{h}[m]$ in the Hamiltonian~\cite{qite_chan20}. For example, the smQITE calculation with qubit domains reduced to $D'=4$, which amounts to a much larger dimension of 3648 for the Pauli basis set, yields an energy over 30 mHa higher for H$_2$O molecule at $R=1$\AA.

\begin{figure}[th!]
	\centering
	\includegraphics[width=0.45\textwidth]{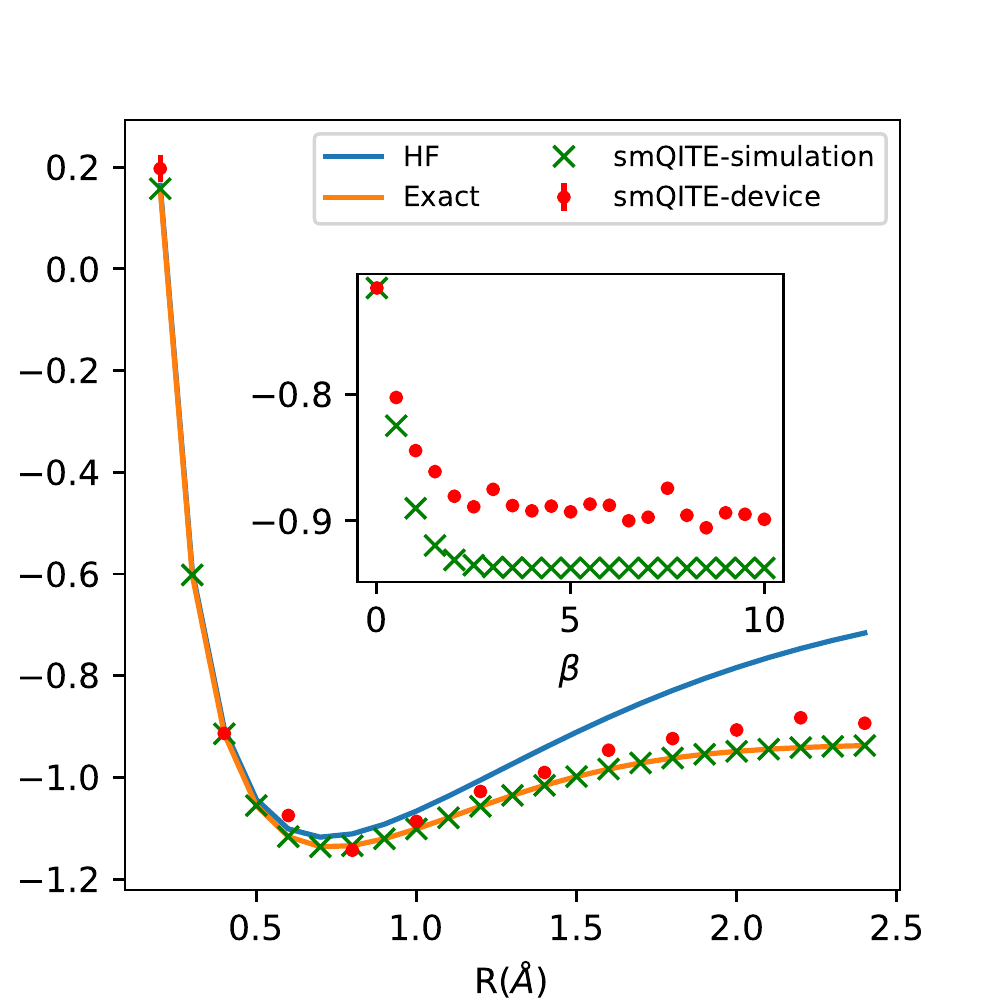}
	\caption{
	\textbf{Demonstration of smQITE calculations of H$_2$ molecule on Rigetti quantum device.} The binding energy curve from smQITE calculations using wavefunction simulator and Rigetti Aspen-4 device are shown, together with the results from ED (FCI) and HF for reference. Inset: smQITE energy evolution as a function of Trotter steps $\beta=n\times\Delta\tau$ for H$_2$ molecule at $R=2.4$\AA\,using wavefunction simulator and real device, with fixed $\Delta\tau=0.5$.
	}
	\label{figure4}
\end{figure}

\subsection{smQITE calculations on quantum devices}
Finally, we benchmark the smQITE calculations on real quantum devices through the quantum cloud service provided by Rigetti. The H$_2$ molecule is chosen as an example for the demonstration. The smQITE calculations with a compact Pauli basis from sUCCSD ansatz are carried out to make an efficient use of quantum resources. As a result, the Pauli basis is composed of a single Pauli term $X_0Y_1$, which is essentially the same of the UCCSD ansatz employed in the literature for VQE calculations of H$_2$ or other similar two-orbital systems~\cite{gqce}. Here $X$ ($Y$) is the x(y)-component of a single qubit Pauli operator. Figure~\ref{figure4} shows smQITE calculations for the total energy of H$_2$ molecule as a function of bond length using wavefunction simulator and Rigetti Aspen-4 device. The wavefunction simulation data overlap with the ED (FCI) results, because the sUCCSD ansatz is exact for this example. The smQITE calculations on real device follow the exact curve quite well, with errors on the order of 10 mHa. The inset plots the energy evolution as a function of Trotter step $\beta=n\times\Delta\tau$ with fixed $\Delta\tau=0.5$ from smQITE calculations of H$_2$ molecule at $R=2.4$\AA\,on wavefunction simulator and the quantum device. Starting from the initial HF state, the smQITE energy decreases as the Trotter step proceeds. The energy points converge to the exact value for smQITE calculations on the wavefunction simulator, which represents the ideal fault-tolerant quantum computer with infinite repeated measurements (shots) of the associated Pauli terms. The smQITE energy from real device calculations drops and fluctuates around a value higher than the exact point, due to the sizable noise in the current real device and finite shots in calculations. The final ten points in the smQITE calculations are used to estimate the mean-values and standard deviations, which are reported in Fig.~\ref{figure4}. The standard deviation is generally within the symbol size. 

The Rigetti 13-qubit Aspen-4 quantum device is used for the above smQITE calculations. Qubits with index 1 and 2 are used to represent the Hamiltonian of H$_2$. The fidelity of the two-qubit gate is about 95\%. At each smQITE step, five different quantum circuits are constructed to measure the expectation values of eight Pauli terms, with some of them measured simultaneously due to mutual commutation. Readout error symmetrization and mitigation, as implemented in the Forest package~\cite{pyquil_1}, have been used to reduce the effects of noise. The readout symmetrization is performed by exhaustively flipping the qubits before the measurements ($2^2=4$ ways for the two-qubit system), and subsequent flipping back the measurement outcomes. As the effect of symmetric measurement error is to scale the expectation value of the Pauli observable by a noise-dependent factor, the error mitigation is to rescale the measured observable expectation value accordingly. The readout symmetrization comes at a price, which effectively introduce $4\times5=20$ quantum circuits at each smQITE step. We use $2^{10}$ shots during the measurement of Pauli terms for each circuit.

\section{Conclusion}
In conclusion, the smQITE algorithm has been developed as a resource-efficient version of QITE, which adapts better to the current and near-term NISQ hardware. Highly accurate results have been demonstrated for the smQITE calculations of the binding and dissociation energy curves of a set of molecules. To simulate molecular Hamiltonian of increasing size, a compact representation of the smQITE unitary evolution operators has been proposed by adopting a variational wavefunction form in VQE calculations. It has been shown that the smQITE calculations converge much faster, and achieve the similar accuracy as VQE with the same variational circuit. Finally, we demonstrate smQITE calculations on a Rigetti quantum device, where the binding energy curve of H$_2$ molecule has been obtained with a reasonable accuracy. Numerical results suggest that the inherent challenge in the non-convex high-dimensional optimization problem of VQE calculations can potentially be addressed by a combination of smQITE and VQE, where the fast-converged smQITE solution can be fed into VQE for further optimizations. 

\section*{Acknowledgements}
This work was supported by the U.S. Department of Energy (DOE), Office of Science, Basic Energy Sciences, Materials Science and Engineering Division. The research was performed at the Ames Laboratory, which is operated for the U.S. DOE by Iowa State University under Contract No. DE-AC02-07CH11358. 

\bibliography{refabbrev, ref}

\end{document}